\documentclass[twocolumn,showpacs,aps,prl]{revtex4}

\usepackage{graphicx}%
\usepackage{dcolumn}
\usepackage{amsmath}
\usepackage{latexsym}

\begin {document}

\title
{
Weighted scale-free network with self-organizing link weight dynamics 
}
\author
{
G. Mukherjee$^{1,2}$ and S. S. Manna$^{1}$
}
\affiliation
{
$^1$Satyendra Nath Bose National Centre for Basic Sciences
    Block-JD, Sector-III, Salt Lake, Kolkata-700098, India \\
$^2$Bidhan Chandra College, Asansol 713304, Dt. Burdwan, West Bengal, India
}

\begin{abstract}

      All crucial features of the recently observed real-world weighted networks
   are obtained in a model where the weight of a link is defined with a single non-linear
   parameter $\alpha$ as $w_{ij}=(s_is_j)^\alpha$, $s_i$ and $s_j$ are the strengths of two end
   nodes of the link and $\alpha$ is a continuously tunable positive parameter.
   In addition the definition of strength as $s_i= \Sigma_j w_{ij}$ results a
   self-organizing link weight dynamics leading to a self-consistent distribution
   of strengths and weights on the network. Using the Barab\'asi-Albert growth dynamics 
   all exponents of the weighted networks which are continuously tunable with $\alpha$
   are obtained. It is conjectured that the weight distribution should be similar in
   any scale-free network.

\end{abstract}

\maketitle

      Representing widely different large interacting systems as networks and identifying
   common generic characteristics in them  made a significant advance in the study of complex
   systems in recent years \cite {Barabasi,Barabasi2,Barabasi3,Barabasi4}. 
   However in most cases, link weights, representing the strength of ties between nodes, are not
   similar and may even be highly heterogeneous. E.g., number of passengers between two airports
  \cite {Guimera,Barrat},
   strength of pair-interaction between two species in ecological system, number of coauthored
   paper of two authors, data traffic in a link of the Internet or volume of trade between two
   countries differ widely \cite {loffredo}. Present availability of empirical data allows to study the 
   weighted networks having weights associated with links. Weighted networks are usually described
   by the adjacency matrix $W$, $ij$-th element $w_{ij}$ of which represents the link weight
   between $i$ and $j$-th node. Recent analysis of the World Airport Network (WAN) data shows
   that the link weights defined by the passenger traffic has power law distribution:
   $P(w)\sim w^{-\gamma_w}$ \cite {Barrat}. 

      Evolution of these weighted networks and the dynamics associated with it is a matter of
  general interest. In a traffic network like WAN when a new airport emerges with few links 
  the passenger traffic associated with these new links affect the traffic
  of the other links also. In turn the effect of the other links also change the traffic
  of the new links and this mutual interaction takes its time to reach a steady state.
  Similar effects are there in many weighted networks.
  In this paper we try to explore this self-organisation of link weights and node strengths 
  with the evolution of the network.

      A large number of models have been proposed to generate different type of networks.
   Most well known among them is the Barab\'asi and Albert (BA) network which grows by sequential 
   addition of new nodes using a linear preferential attachment probability. These are unweighted 
   networks where each link has the same weight.

      Strength of a node is measured by the total amount of weight supported by the node:
   $s_i= \Sigma_j w_{ij}$. The WAN data shows non-linear scaling between
   average strength of a node with its degree: $ s(k) \sim k^\beta $ with $ \beta \sim 1.5$.
   On the other hand average link weight again scales with the product of the degrees of 
   the two end-nodes : $\langle w_{ij} \rangle \sim (k_ik_j)^\theta$, where $\theta =0.5$
   for WAN and $0.8$
   for E. Coli metabolic networks where link weights represents the optimal metabolic
   fluxes \cite {Wang, Bianconi, Goh}.

      Barrat, Barth\'elemy and Vespignani (BBV) proposed a model\cite { Barrat1}
   for the evolution of the weighted network similar to the BA model. A new node $n$ selects
   a node $i$ of the network with a probability proportional to its strength:
   $\Pi_{n\rightarrow i} \sim s_i$. In addition, $\delta$ amount of weight is 
   distributed proportionally among the already existing links of the target node, hence with
   the topological evolution of the network link weights also evolve. BBV model produces 
   power-law distribution of link weights, node degree and strength scales linearly and
   both exhibit power-law with same exponent. 
         
      Even on an unweighted network one can assign link weights as $ w_{ij} \sim (k_ik_j)^\theta $.
   This is a static assignment for a given structure. The idea is, large degree nodes process
   large amount of traffic and in the level of links, traffic through a link grows with the degree
   of its two end nodes. Using this idea we propose a model where the weight of a link depends
   jointly on the strengths of the end nodes which modifies the strengths of the nodes and which
   recursively modifies the link weights and so on until all link weights and strengths converges
   to a specific set of fixed values. Apart from power law degree distribution we
   find power-law distributed probability of nodal strengths and link weights and the exponents
   decays with the model parameter $\alpha$. The other crucial empirical observation for 
   weighted networks (a) degree and strength of a node scales in a non-linear way and (b)
   a sub-linear variation of link weight $w_{ij}$ with the product of the end-node degrees 
   is also found to be present and the respective exponents $\beta$ and $\theta$ varies with $\alpha$.    

      In our model, the growth of the network structure having weights associated with every link
   is executed by two mutually independent processes as follows:
   (i) The network is a BA network and is grown using the simple algorithm described 
       in \cite {mukherjee}. Starting from a small fully connected $m$-clique, at every
       time step one link is selected with uniform probability and the new node is connected
       to one of the two end nodes with probability 1/2.
  (ii) After every node is added to the network, the weights of all links are re-determined
       by a self-consistent iterative process such that the weight $w_{ij}$ of the link between 
       the $i$-th and the $j$-th nodes depends on the nodal strengths $s_i$ and $s_j$ as:
\begin {equation}
s_i= \Sigma_j w_{ij} \hspace*{1.0 cm} {\rm and} \hspace*{1.0 cm} 
w_{ij} \sim (s_is_j)^\alpha 
\end {equation}
   where $\alpha$ is a continuously tunable parameter. Indeed the dependence of link weight on the
   strengths of the end nodes has been observed in the International Trade Network with $\alpha \approx 1.33$
   \cite {ITN}.

\begin{figure}[top]
\begin{center}
\includegraphics[width=6.5cm]{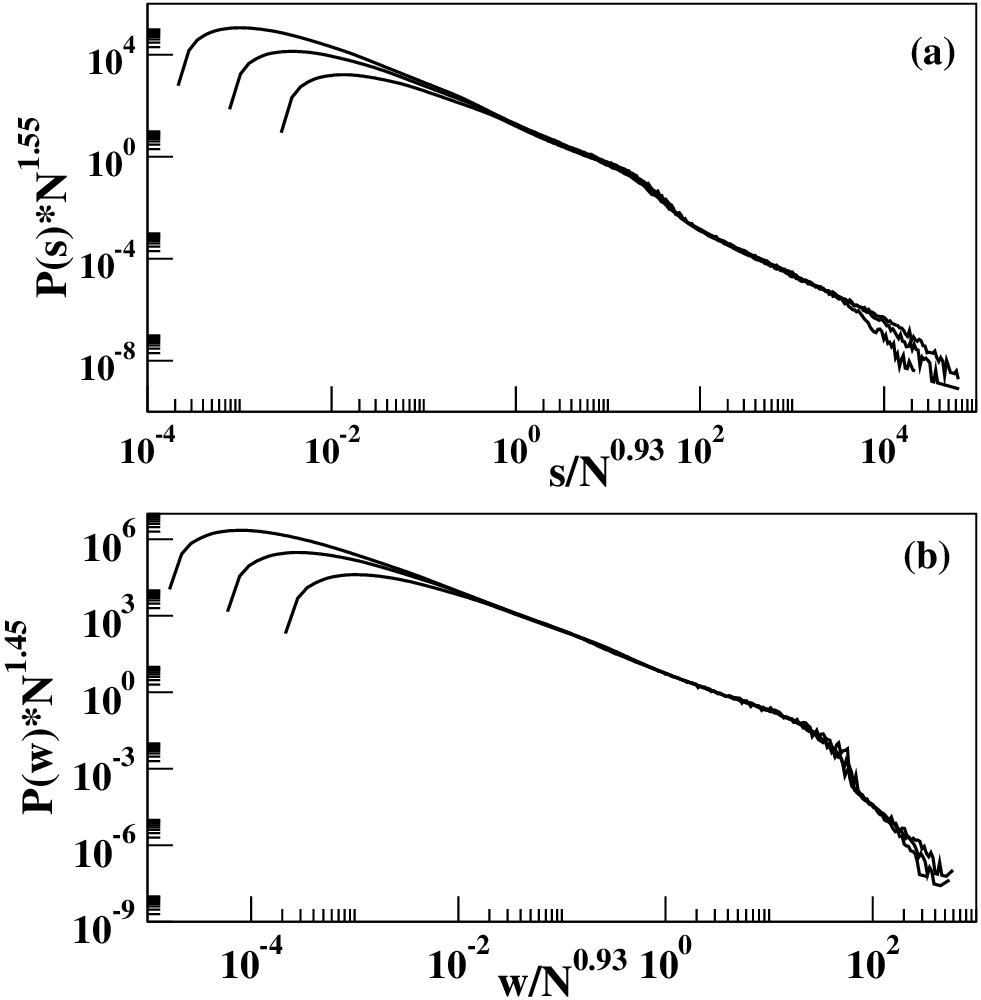}
\end{center}
\caption{Scaling plot of probability distribution for (a) node strengths for $N = 2^{12},
2^{14}, 2^{16}$ for $\alpha = 0.40$, (b) link weights for $\alpha = 0.40$, for $N = 2^{16},
2^{18}, 2^{20}$ . 
}
\end{figure}

      At every time step a new node $i$ and a new link $ij$ with unit weight is added to the network
   increasing the degree as well as the strength of the node $j$ attached. As a result the weights of
   the other links meeting 
   at $j$ increases as well, which in turn enhance the strengths of the neighbors of $j$. In this way the
   effect of inserting a new node spreads throughout the network. One can imagine two widely apart time scales
   are involved in this process: the slow time scale characterize the network growth process whereas the
   fast time scale determines the re-organization of the link weights and strengths.

      Given a connected network, the distribution of weights over all links and the distribution of
   strengths over all nodes consistent with Eqn. (1) can be calculated in a self-consistent 
   iterative process. Let $s^n_i$ and $w^n_{ij}$ denote the $n$-th update of the strength of the $i$-th
   node and the weight of the $ij$-th link. In the self-consistent procedure, initially all links are 
   assigned unit weights, $w^0_{ij}$ = 1. The nodal strengths are then determined as: $s^0_i=\Sigma_j w^0_{ij}$. 
   In the first stage of iteration process, the weights are determined as: $w^1_{ij}= (s^0_i. s^0_j)^\alpha$ 
   and consequently the strengths as: $s^1_i = \Sigma_j w^1_{ij}$, and so on. More specifically,
\begin {eqnarray}
w^0_{ij} & = & 1 \hspace*{1.05 cm}                    \rightarrow s^0_i=\Sigma_j w^0_{ij}=k_i ; \nonumber \\
w^1_{ij} & = & (s^0_i. s^0_j)^\alpha \rightarrow s^1_i=(s^0_i)^\alpha\Sigma_j(s^0_j)^\alpha ; \nonumber \\
w^2_{ij} & = & (s^1_i. s^1_j)^\alpha \rightarrow s^2_i=(s^0_i)^{\alpha^2}\Sigma_j[(s^0_j)^{\alpha^2}(\Sigma_j(s^0_j)^\alpha)^\alpha \nonumber \\
         &   & \hspace*{5.00 cm}                         (\Sigma_k(s^0_k)^\alpha)^\alpha] ... \nonumber
\end {eqnarray}
\begin{figure}[top]
\begin{center}
\includegraphics[width=6.5cm]{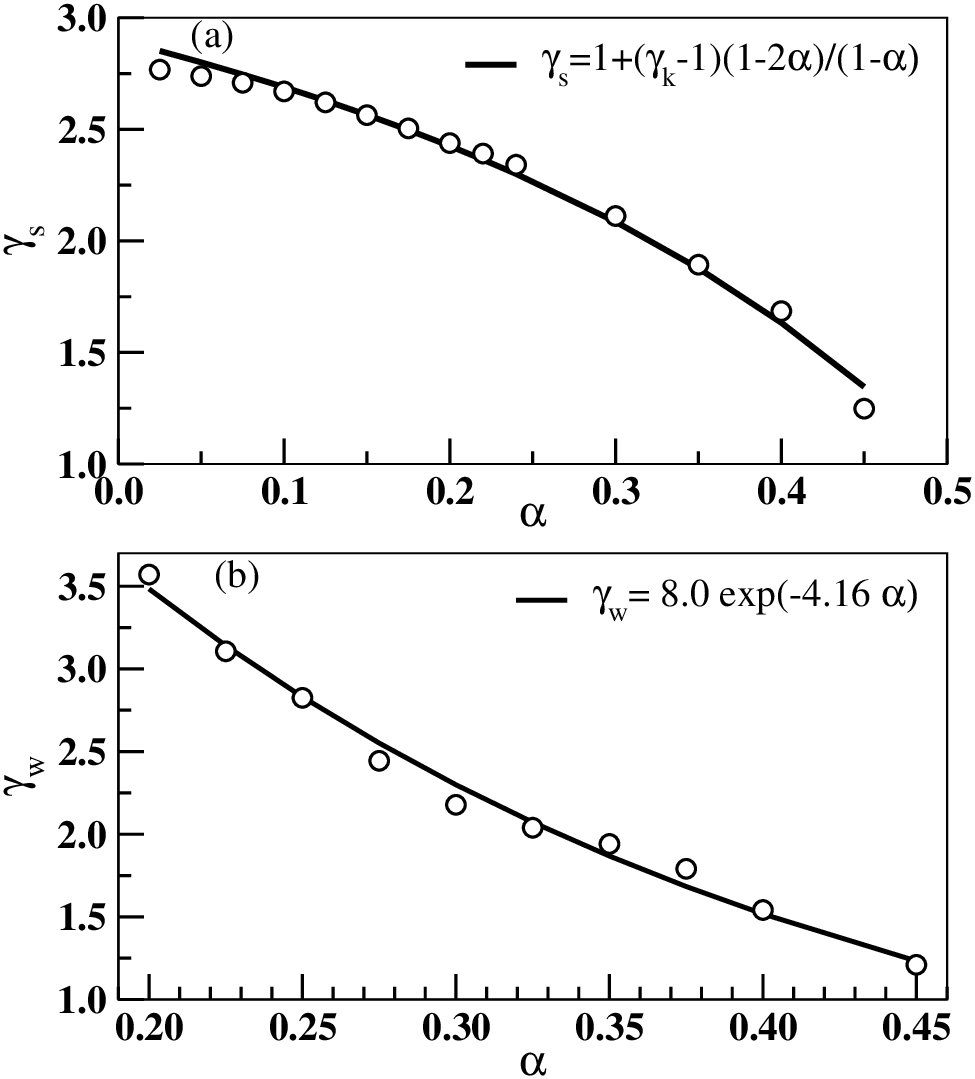}
\end{center}
\caption{(a) Variation of strength distribution exponent $\gamma_s$ with parameter $\alpha$ and
(b) variation of link weight distribution exponent $\gamma_w$ with $\alpha$ for $N = 2^{16}$ in
 both cases. Circles represents the data points and  the continuous lines are the theoretical
 prediction of eqn.(6) and the exponential fitting function, drawn for reference. 
}
\end{figure}
   Under this successive iteration process, the distribution of link weights as well as nodal strengths 
   gradually converge to fixed values when further iterations do not change their values any more. In
   practice we associate a tolerance $\delta = 10^{-6}$ for the maximal difference of weights in a single
   link between two successive iterations. As described above that in $s^2_i$ sums run up to the
   second neighbours which implies that estimation of $s^n_i$ should have sums running     
   up to the $n$-th neighbours to consider the contribution up to $n$-th shell
   neighbours. Since BA network has the small-world properties of small diameters (diameter ${\cal D}(N)$
   of the network grows logarithmically with system size $N$) therefore from an arbitrary node the whole
   network is reached within a few steps.

      Let us first calculate how the mean nodal strength $\langle s(k) \rangle$ of the $k$-degree nodes
   depends on the degree $k$ as a function of the tuning parameter $\alpha$. Let an arbitrary node $i$
   have degree $k_i$ and the mean degree of its nearest neighbours be $k_{nn}$. Following the self-consistent
   iterative procedure, the strength $s_i$ of the node $i$ after $n$-th iteration is:
\begin {eqnarray} 
s_i(n) & = & k^{1+\alpha(1+2\alpha(1+2\alpha(.......n-th (1+\alpha))))} \nonumber \\    
       &   & k_n^{\alpha(1+2\alpha(1+2\alpha(1+2\alpha(...n-th))))}k_{nn}^{2(n-2)\alpha^n} \nonumber \\
s_i(n) & = & k^{M(n)} k_n^{D(n)}k_{nn}^{2(n-2)\alpha^n}   
\end {eqnarray}
      where $n$ is an arbitrarily large integer. Now for a steady state value of $s_i(n)$ both the series
      $M(n)$ and $D(n)$ must converge and reach a steady state value for any arbitrarily large $n$. 
      The convergence condition gives:
\begin {eqnarray} 
 M(n) & = & 1+\alpha(1+2\alpha(1+2\alpha(.......n-th (1+\alpha))))\nonumber \\  
      & = & 1+\frac {\alpha}{1-2\alpha}
\end {eqnarray}
 and    
\begin {eqnarray} 
 D(n) & = & \alpha(1+2\alpha(1+2\alpha(1+2\alpha(...n-th))))\nonumber \\  
      & = & \frac {\alpha}{1-2\alpha}
\end {eqnarray}
      For finite steady state value of $M(n)$ and $D(n)$ both of the above equation leads to the
      condition:   $\alpha<0.50$, beyond which link weights and node strengths will diverge.

\begin{figure}[top]
\begin{center}
\includegraphics[width=6.5cm]{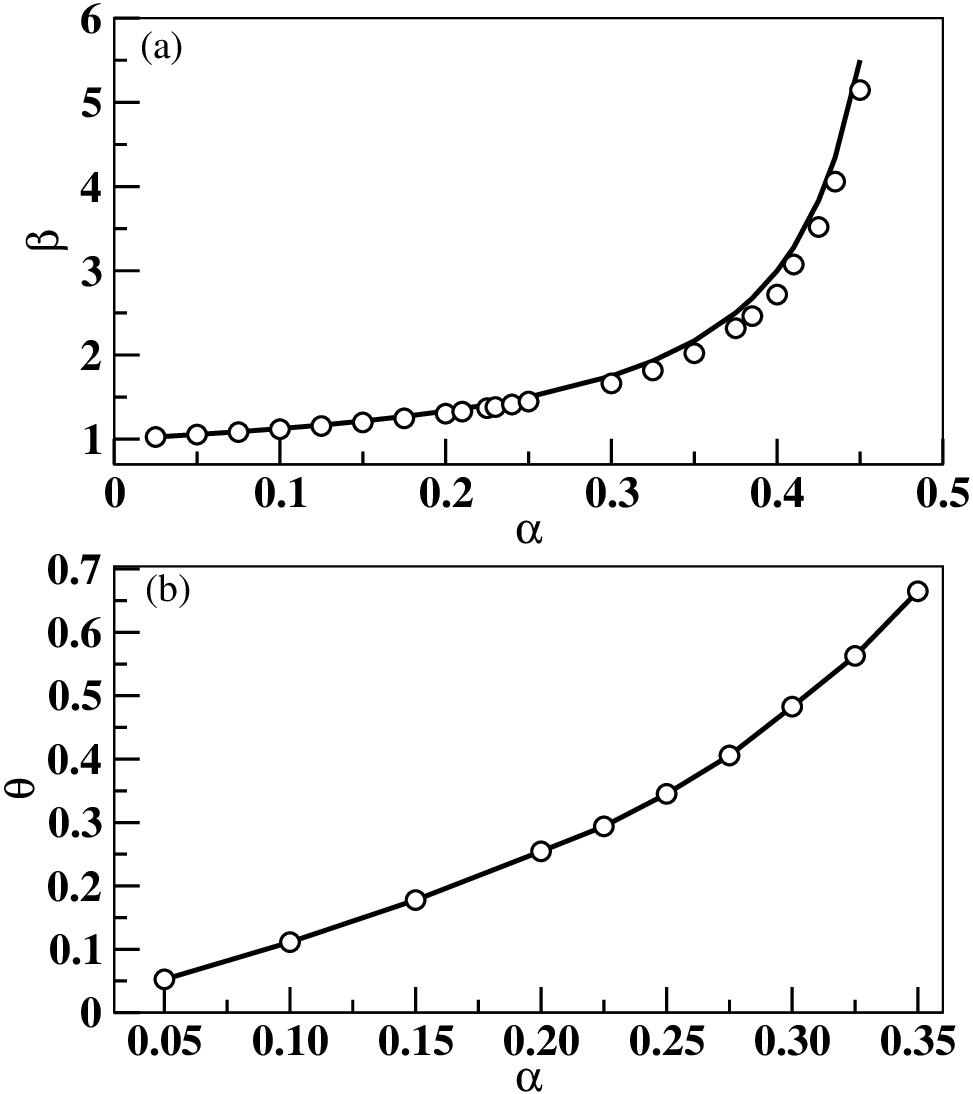}
\end{center}
\caption{ (a) Variation of strength degree correlation exponent $\beta$ with $\alpha$, open circles
 are data points and the solid line is the theoretical prediction of eqn.(5). (b) Plot of link weight
 degree correlation exponent $\theta$ with $\alpha$, $N = 2^{16}$, solid line is the plot of the fitting
 function $\theta = a + b \alpha^c$ .
}
\end{figure}

      Now  for strength-degree dependence $s \propto k^\beta$ we assume $k_n$ has no
    significant dependence on $k$, which is true for BA topology, hence
\begin{equation}
\beta \approx M(n) = 1+\frac {\alpha}{1-2\alpha}
\end{equation}         
To find the exponent $\gamma_s$ of the probability distribution of
   node strengths $P(s) \sim s^{-\gamma_s}$ we use the general relation
   $\beta=\frac{\gamma_k-1}{\gamma_s-1}$. Which gives,
\begin{equation}
\gamma_s= 1+\frac {(\gamma_k-1)(1-2\alpha)}{1-\alpha} 
\end{equation}         

      Probability distribution of strengths
   shows power law decay: $P(s) \sim s^{-\gamma_s}$ over several decades in Fig. 1(a).
   We find that a `knee'-like
   step is developed in the distribution near $\alpha \sim 0.25$ and remains up to
   $\alpha \sim 0.45$. With the increase of $\alpha$ from $0.25$ the position of the knee
   shifted to wards the higher value of strength and become more prominent. In Fig.1(a)
   we have shown the scaling plot of $P(s)\sim s$ at $\alpha = 0.40$, where the 'knee'
   is quite prominent. The scaling function :
\begin {equation}
P(s) \sim N^{\lambda} {\cal G}(s/N^{\mu})
\end {equation}
   where $\lambda \approx 1.55 $ and $\mu \approx 0.93$ are estimated giving $\gamma_s = 
   \lambda / \mu \approx 1.67$. This value is consistent with the value found from
   slope measurement of the $\log P(s)$ vs $ \log(s)$ plot ($\approx 1.68$).

\begin{figure}[top]
\begin{center}
\includegraphics[width=6.5cm]{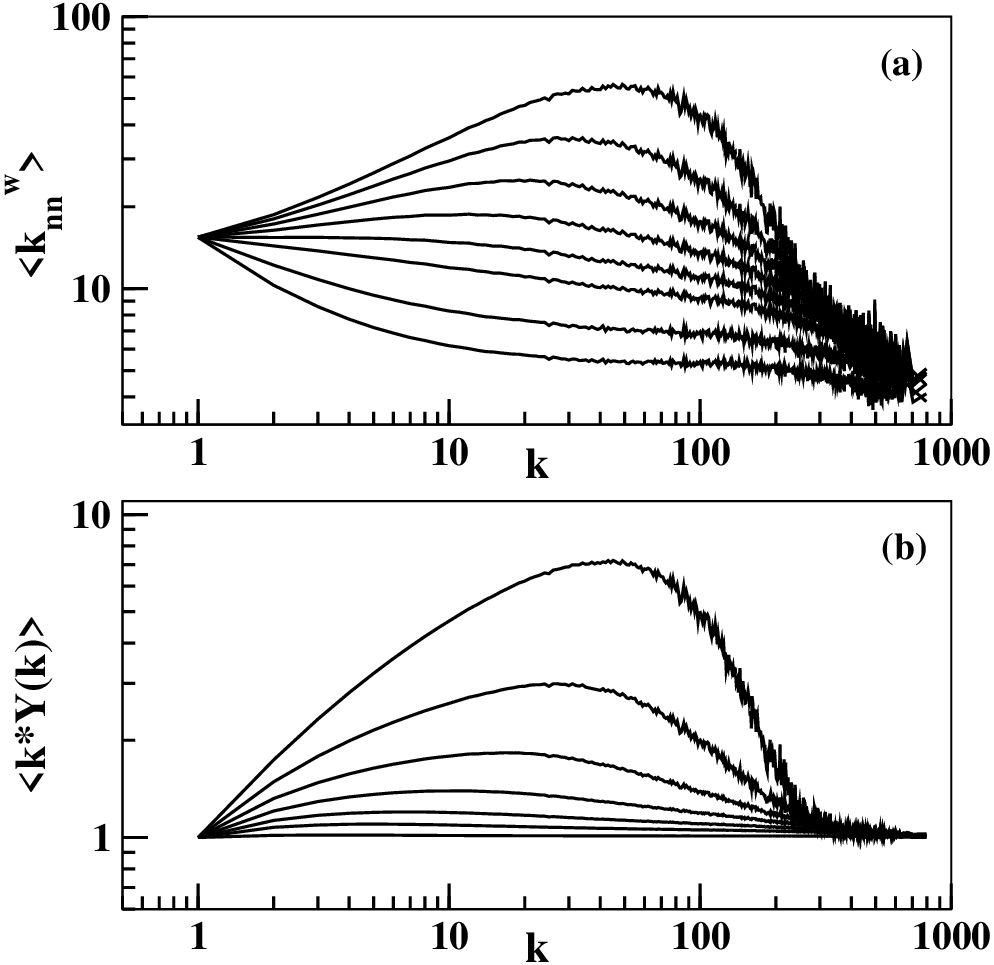}
\end{center}
\caption{Plot of (a) Weighted average neighbour degree $\langle k^w_{nn}\rangle $ with $k$ 
for $\alpha =$0.00, 0.10, 0.20, 0.25, 0.30, 0.35, 0.40, 0.45 from bottom to top,
 serially (b) average disparity measure $<k*Y(k)>$ with $k$ for the same $\alpha$ values
from bottom, to top, $N = 2^{14}$.
}
\end{figure}

       The probability distribution of the link weights again shows power-law variation:
   $P(w) \sim w^{-\gamma_w}$. For very small values of $\alpha \approx 0.10$ link weights
   are very small, even for a system of size $N = 10^6$ maximum value of $w$ is around $10$
   and power-law region is not well-defined. In Fig. 1(b) scaling plot of $P(w)$ is shown
   for $\alpha=0.40$ and with $N=2^{16}, 2^{18}, 2^{20}$. Weight exponent $\gamma_w$, found from 
   scaling analysis ($= 1.45/0.93\approx 1.56$) is close to the value found from slope measurement
   of $\log P(w)$ vs $\log w$ plot, ($\approx 1.55$).  

      In fig. 2(a) variation of strength exponent $\gamma_s$ with $\alpha$ is plotted. Open circles
   are experimental points and the solid line is the theoretical prediction. At larger
   values of $\alpha$ maximum value of strength will increase and the probability of occurrence
   of large strength nodes should be increased.
   Drop in the exponent $\gamma_s$ of the power-law decay represents this trend.
   The data points and the expression of $\gamma_s$ both indicates the value 
   of $\gamma_s$ at $\alpha\approx 0.50$ as $1.0$. From the measurement of the exponent $\beta$ 
   also it was found that the value of $\beta$ is asymptotically large as $\alpha\rightarrow 0.50$.
      In fig.2(b) variation of link weight distribution exponent $\gamma_w$ with $\alpha$
   is shown. The initial high $\gamma_w$ value shows that indeed for small $\alpha$ the
   distribution decays very fast. Data points are compared with a fitting function: 
   $ \gamma_w = a.\exp(-b.\alpha)$, $a = 7.98, b = 4.16$. This exponential decay of $\gamma_w$
   again signifies that link weights grow fast with $\alpha$ in general but specially
   the growth of large links are much faster.

      Fig. 3(a) shows strength-degree exponent $\beta$ increases with $\alpha$.
   While $\beta= 1$ implies that link weights are uncorrelated, for $\beta>1$ 
   large weights prefer to join the large degree nodes. Increase 
   in the value of $\beta$ means more aggregation of large weight links on large degree nodes.
   That signifies that $\alpha$ systematically
   increases the preference of large weights to converge on high degree hubs. The other
   factor is that with the increase of $\alpha$ degree values of the nodes remains unchanged
   but links of very large weights and hence very large node strengths appear and contributes 
   to $\beta$. The plot show a steep rise in $\beta$ in the region $\alpha> 0.40$.
       In fig. 3(b) link weight-end degree exponent $\theta$ is plotted with $\alpha$. 
   Exponent $\theta$ found to grow with $\alpha$. It is expected here as with $\alpha$ link 
   weights increases but the degree distribution and hence the degree of any two
   end nodes remains unchanged. Plot is fitted with a function of the form:
   $\theta = a + b \alpha^c$, with $a = 0.06, b = 5.23, c = 2.07$ where $a\approx0.00$
   is more appropriate as it corresponds to $\alpha=0.00$ value.

      The degree-degree correlation $\langle k_{nn} \rangle$ is \cite {Barrat}:
\begin {equation}
k_{nn,i} = \frac {1}{k_i}\Sigma_jk_j
\end {equation}
    where the sum $j$ runs over the nearest neighbours of $i$.
    Now when averaged over the nodes according to degree $k$ we find $ k_{nn}(k)$ which is
    related to the conditional probability $P(k_1|k)$ as that a node of
   degree $k$ is connected to a node of degree $k_1$: 
\begin {equation}
k_{nn}(k) = \frac {1}{N_k}\Sigma_{k_i=k} k_{nn,i}=\Sigma_{k_1} k_1 P(k_1|k)
\end {equation}
    If there is no degree-degree correlation $P(k_1|k)$ is a function of $k_1$ only and
    $k_{nn}(k)$ is a constant. If $k_{nn}(k)$ increases (decreases) with $k$, the network 
    is said to be `assortative' (`disassortative').
   In weighted networks
   weighted average nearest neighbour degree is defined as follows:    
\begin {equation}
k_{nn,i}^w = \frac {1}{s_i}\Sigma_j w_{ij}k_j
\end {equation}
    and similarly the $k_{nn}^w(k)$.

   It is known that in BA model there is no significant degree correlation. This has been
        observed in our calculation also that in Fig. 4(a) for $\alpha=0$, the curve is nearly horizontal
        for a large portion of the whole range of variation of the degree $k$.
        Therefore the underlying BA network structure cannot be responsible for additional
        correlation in $k^w_{nn}(k)$ in Fig. 4(a) for $\alpha > 0$ and it must be due to the
        self-consistent procedure with which the links are updated.

        Our understanding is that the assignment
        of link weights as $w_{ij} \propto (s_i s_j)^{\alpha}$ induces this assortative nature. As
        initially $s_i=k_i$, weights of the links between large degree nodes become large
        and that produces the assortative nature in the weighted degree correlation
        as     $k_{nn,i}^w = \frac {1}{s_i}\Sigma_j w_{ij}k_j$.

      The nodal disparity or inhomogeneity is $Y_i=\Sigma_j (w_{ij}/s_i)^2$. When all the 
   links contributed equally $Y_i \sim 1/k_i$ and when one link gives the major contribution 
   and others are not significant then $Y_i \sim 1$. Network disparity $Y(k)$ is the 
   nodal disparity averaged over all nodes of degree $k$, shown in Fig. 4(b).
   For small values of $\alpha$ the plot
   is almost horizontal and the value is $1$, indicating homogeneity of links for 
   different node degrees. But for $\alpha>0.35 $ there is inhomogeneity within a certain
   region of $k$ value. First it starts growing then decreases to $1$ again for hub nodes.
   This signifies that for a broad region of $k$ values nodes have highly heterogeneous
   link weights but the links of the hub nodes are in general have very large weights and
   not heterogeneous in nature, that again is the effect of large link weights aggregation
   on hubs.   
  
      To summarize, we have shown that starting from an unweighted network one can have 
   non-trivial distribution of link weights as well as nodal strengths using only the
   definition of strength as well as the relation describing the non-linear dependence of
   link weights on the strengths of end nodes. The distribution of weights defines the
   strengths which in turn modifies the link weights - as a result link weights and nodal
   strengths self-organise themselves to a steady-state distribution through a recursive 
   process. Study of this process on the BA network shows rich properties like power-law
   distributed probability distribution of nodal strengths as well as link weights, with
   associated exponents varying continuously with the model parameter $\alpha$. 
   The model exhibits all non-trivial features of real-world weighted networks, e.g., 
   non-linear strength-degree relation, correlation between link weights and the product 
   of degrees of the end-nodes etc. We conjecture that similar non-trivial weight and
   strength distributions can be obtained in general for any scale-free network
   and therefore may be useful to study different real-world weighted networks. 

   GM thankfully acknowledged facilities at S. N. Bose National Centre 
   for Basic Sciences. 

\leftline {Electronic Address: manna@boson.bose.res.in}

\end {document}